\documentclass{ae}[times]  
\usepackage{graphicx}
\usepackage{txfonts}
\usepackage{hyperref}
\begin{document} 

   \title{{\em \Large{Very Important\\Letter to the Editor}}\vspace{0.6cm}\\ Follow the Index: A new proposal}
\titlerunning{The b-index}
\authorrunning{Henri et al.}

   \author{Henri M.J. Boffin
          \inst{1}\fnmsep\thanks{As for all his papers, this author is sole responsible for its content, which does not represent in any way or another, not even when seen through a telescope, the views of his employer, real or supposed.}
                 }

   \institute{$^1$ Extraterrestrial Institute for the Advancement of Earth (EIAE),
            Secret place, Planet Earth, Solar System  
                           }

   \date{Received March 30, 2022; accepted March 31, 2022}

  \abstract
   {Despite all its well-known flaws and calls for its dismissal, the notorious $h$-index is still used in many instances when awarding grants, or promoting and hiring scientists. To address this, I set out to devise a better index, with the twofold aim of taking into account the authors' respective contributions and considerably reducing the pollution of the scientific literature. Finally, I present a strategy that is guaranteed to be best for all researchers. 
 
   \vspace{0.25cm}}
   \maketitle
\begin{flushright} {\it If your result needs a statistician\\ then you should design a better experiment. }\\  --- Ernest Rutherford
\end{flushright}

\section{Defined by a number}

{\it ``I am not a number! I'm a free man,''} says the Prisoner in the psychedelically experimental TV series of the same name. This is a sentence that surely every scientist must have uttered at one point or another of their career, especially when asked to provide their CVs and make sure they clearly indicate their $h$-index and the journal impact factor (JIF\footnote{Not to be confused with the peanut butter or the cleaning product of the same name!}) of each of their publications. It is indeed quite remarkable, for lack of a better word, that scientists, whose main role is to conduct {\it a systematic enterprise that builds and organizes knowledge in the form of testable explanations and predictions about the universe \footnote{\url{https://en.wikipedia.org/wiki/Science}}''} and thereby follow a well-defined method, scientists, thus, are often just defined by a number -- their $h$-index. As if we were to only characterise our friends and family by their IQ! 

It is true that any scientific result is only worth anything if it is communicated -- first and foremost with your peers. This is done in various forms, such as informal discussions over coffee, talks at conferences, publications in scientific journals, and writing grant and job applications. Still, a researcher's career is often only judged by their publications. Initially, this was done by looking at how many refereed papers someone would have, in which journal, and how many citations did they collect. But very quickly, even this, was too tedious, and it became easier to just decide who to fund, who to employ or who to promote, based on a number only -- the $h$-index.

If this is better than using the JIF to define someone -- something not very different from judging a person by the brand of the shoes they wear -- it is quite appalling that the intricacies of a person's life and their achievements should be summarised by one number. Especially as this procedure is not even hidden, and despite so many calls for its dismissal, the $h$-index has become one of the most widely used metrics to measure the productivity and impact of scientists.

\section{The non hilarious index}
The $h$-index\footnote{Not to be mistaken for the Heat-index; see \url{https://xkcd.com/2026/}}, where strangely enough the $h$ does not stand for ``hilarious'', is the largest number $h$ such that at least $h$ publications have been cited at least $h$ times each \citep{2005PNAS..10216569H}. When J.E. Hirsch introduced the $h$-index in 2005, his hope was that this is {\it ``a useful index to characterize the scientific output of a researcher,''} and then went on to suggest {\it ``that for faculty at major research universities, $h \approx 12$ might be a typical value for advancement to tenure (associate professor) and that $h \approx 18$ might be a typical value for advancement to full professor.''} This suggestion was, rather unfortunately, promptly endorsed by many science administrators, leading to the damages that we know. Fifteen years after introducing the index, the same author \citep{2020arXiv200109496H} expressed some regrets: {\it ``I have now come to believe that it can also fail spectacularly and have severe unintended negative consequences. I can understand how the sorcerer’s apprentice must have felt.''} We, definitively, still feel it!

Perhaps one should note that if a value of 18 seems enough for becoming a full professor, one must admit that the dream of every researcher is to reach a $h$-index of 42, and thereby find the 
{``Answer to the Ultimate Question of Life, the Universe, and Everything''} \citep{DA42}.

Quite remarkably, \cite{2014arXiv1402.4357Y} has shown that the $h$-index can be reinterpreted as the rule of thumb $h=0.54 \sqrt{C_t}$, where $C_t$ is the total number of citations. An $h$-index of 42 then corresponds to the square of 77.7777777777... citations. Coincidence, you say? I wouldn't be so sure!

\begin{figure}
    \begin{center}
        \includegraphics[width=9cm]{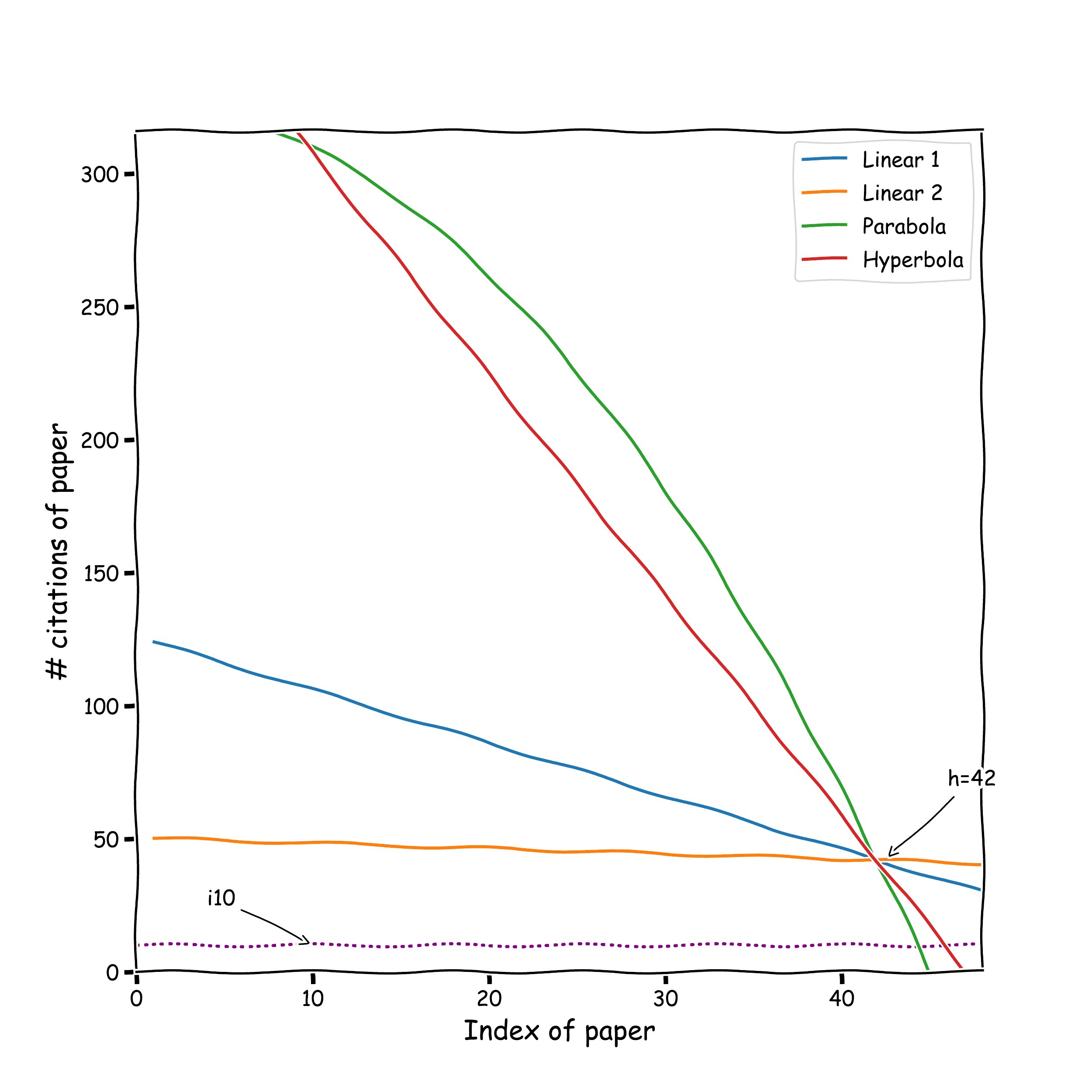}
    \caption{The unbearable lightness of the $h$-index. Several distributions of citations are shown, all characterised by the same value of the  $h$-index, even though they are clearly very different, with differing numbers of total citations and values of the $i10$-index.}
    \label{fig:h}
    	\end{center}
\end{figure}

\begin{table}
\caption{Examples of functional forms shown in Fig.~\ref{fig:h} and their related total citations and $i10$-index.}
\label{tab:h}
\begin{tabular}{lll}
\hline
Function & Total citations & $i$10 \\
\hline
Linear 1 &  3906 &  58 \\
Linear 2 &  3999 &  201 \\
Parabola & 9\,611 &  44 \\
Hyperbola & 9\,003 &  45 \\
\hline
\end{tabular}
\end{table}

\section{Appraisal}
According to its advocates, the $h$-index is a useful and simple way to compare researchers and allow to distinguish between a scientist who has many lower impact publications and someone who has fewer publications, but with a higher impact. However, as every researcher knows, the devil is in the detail. A single number cannot tell the whole story. 

Figure~\ref{fig:h} shows examples of distributions of citations for some hypothetical scientists, all characterised by the same value of the $h$-index, $h=42$ (what else!). It can be seen that they correspond to quite different distributions, with some having produced many papers with moderate citations, and others having perhaps less papers, but a few with quite a remarkable number of citations. Thus, as shown in Table~\ref{tab:h}, this leads to a different number of total citations -- the rule of thumb predicting an intermediate number of 6\,049 -- and a different number of papers with ten citations (the $i10$-index!). Surely, to promote or hire a scientist, we would like to have more than just the $h$-index. 

There are also many other problems with this notorious number. It wouldn't be difficult, for example, for some unethical scientists to get together and agree to often cite each other’s work to artificially increase their index -- a practise not very different from major chains agreeing on retails prices, except that in the latter case, the legislator may intervene.

Perhaps more annoying is that the $h$-index doesn't tell us anything about whether the researcher was the sole or main author of the papers in Fig.~\ref{fig:h}, or if they were just one among hundreds of co-authors. Thus, a sure way to boost your index is to be part of a huge collaboration that makes plenty of self-citations! 

Finally, {\it ``a big drawback of the h-index is that it tells you nothing about the science or ideas behind a researcher’s achievements''}\footnote{\url{https://www.enago.com/academy/drawbacks-of-h-index/}}.

We are thus going back to kindergarten strategy: ``my $h$-index is bigger than yours!\footnote{See for example, \url{https://www.linkedin.com/pulse/my-h-index-bigger-than-yours-humourous-review-2019-isenring/}.}''
Some websites\footnote{\url{https://www.enago.com/academy/how-to-successfully-boost-your-h-index/}} even propose ways to boost your $h$-index!

Thus, its use, especially for hiring and promoting scientists, rends my heart right out of my chest.

\section{A new index}
The previous discussion has clearly demonstrated that the $h$-index is a very poor representation of the output of a researcher and certainly of its quality. There is thus the need for a more sophisticated index. Not being afraid of herculean tasks, I therefore set to devise one. 

I first wanted to call this the ``admirably moderated bibliographic index'' or  ``ambidex'', but it didn't feel right, so I left it. 
The z-index was also already taken and as my goal is to make things easier, not more confusing, I didn't want to grizzle over it. And similar to the $h$-index, the $k$-index (for Kardashian\footnote{\url{https://genomebiology.biomedcentral.com/articles/10.1186/s13059-014-0424-0}}) and the $n$-index (for Narcissist\footnote{\url{http://ecoevoevoeco.blogspot.com/2016/01/a-narcissist-index-n-index-for-academics.html}}) have already been taken. Thus, for lack of a better name, I decided to devise the $b$-index, where $b$ stands of course for... {\it better}. 

The first step to take is to ensure that the index is fairer and takes into account the number of authors on a paper. A simplified version of such a scheme was already proposed by \cite{2014arXiv1408.3881A}, without much success it seems.
The basic idea is that the citations are distributed among the authors according to their relative contributions. However, using a simple scheme is, here again, doomed for failure, as life isn't easy and contrarily to what an actress said, it ain't simple neither, so trying to reproduce it requires a dedicated strategy.

Let us thus consider a paper that has $C$ citations, after having removed all the self-citations, and $N$ authors. Even in the most perfect democracy, these authors aren't all equivalent, however, and we need to make some distinctions: thus, we have the main authors ($A_m$), the secondary ones ($A_s$), and the alphabetically ones ($A_\alpha$). The main authors consist usually of just the first author, and most credit (and citations) should go to them. However, some papers are sometimes written by 2 main authors (as referred in, for example, a note), or sometimes they are written by students, and in this case, it perhaps makes sense to also give full credit to the supervisor(s). The main authors ($m=1,N_m$, with $N_m$ the number of main authors) are each given a non-normalised score, $S_m$, of 1. To make things even more fun and as a consequence of some historical idiosyncrasy, in some special cases, one of the main authors may be the last author in the list of authors!

The secondary authors (but all, very important persons) are those that come after the main authors, as having made some contributions to the paper and the work reported. They shouldn't be listed in alphabetical order. These $N_s$ authors will get a score, $S_s=1/i,$ where $i$ is their position in the author list ($i=N_m+1,N_m+N_s$).

Then, in papers with large author lists, there are the $N_\alpha$ authors who are listed in alphabetical order, after the secondary authors. These must be some reasons why they are listed on the papers and it is not my habit to criticise other people's choices. They should each get a fraction of the score associated to the position $N_m+N_s+1$, that is, they get a score $S_\alpha=\frac{1}{N_m+N_s+1}\frac{1}{N_\alpha}$.

The final score is thus:
$$ S = S_m + \sum_{i=N_m+1}^{N_m+N_s+1}{\frac{1}{i}}, 
$$
and each author $k$ (=$m,s,\alpha$) is given a fraction of the total number of citations according to $$ C_k = C \frac{S_k}{S}.
$$ 
For a given author, the $b$-index is the total of their $C_k$ over their career, to avoid falling into the pitfalls demonstrated by Fig.~\ref{fig:h}. 

To make it easier to calculate the $b$-index, it is then preferable if authors start to be identified in the papers, with a superscript $m,s,$ or $\alpha$. This is a small price to pay for having a system that is so much better.

One should notice that, similarly to the $h$-index, the $b$-index will be a function of the scientific age of a researcher -- this is most acceptable, as humans also get older, and should be taken into account when dealing with scientists and only compare apples with apples\footnote{Or oranges with oranges; or any fruits you prefer!}. We should, however, avoid simply dividing with the time between the first and last paper (which defines the $m$-index) as this, again, makes abstraction of any complicated background someone may have, nor does it consider the fraction of time one can spend on research. 

\begin{figure}
    \begin{center}
        \includegraphics[width=9cm]{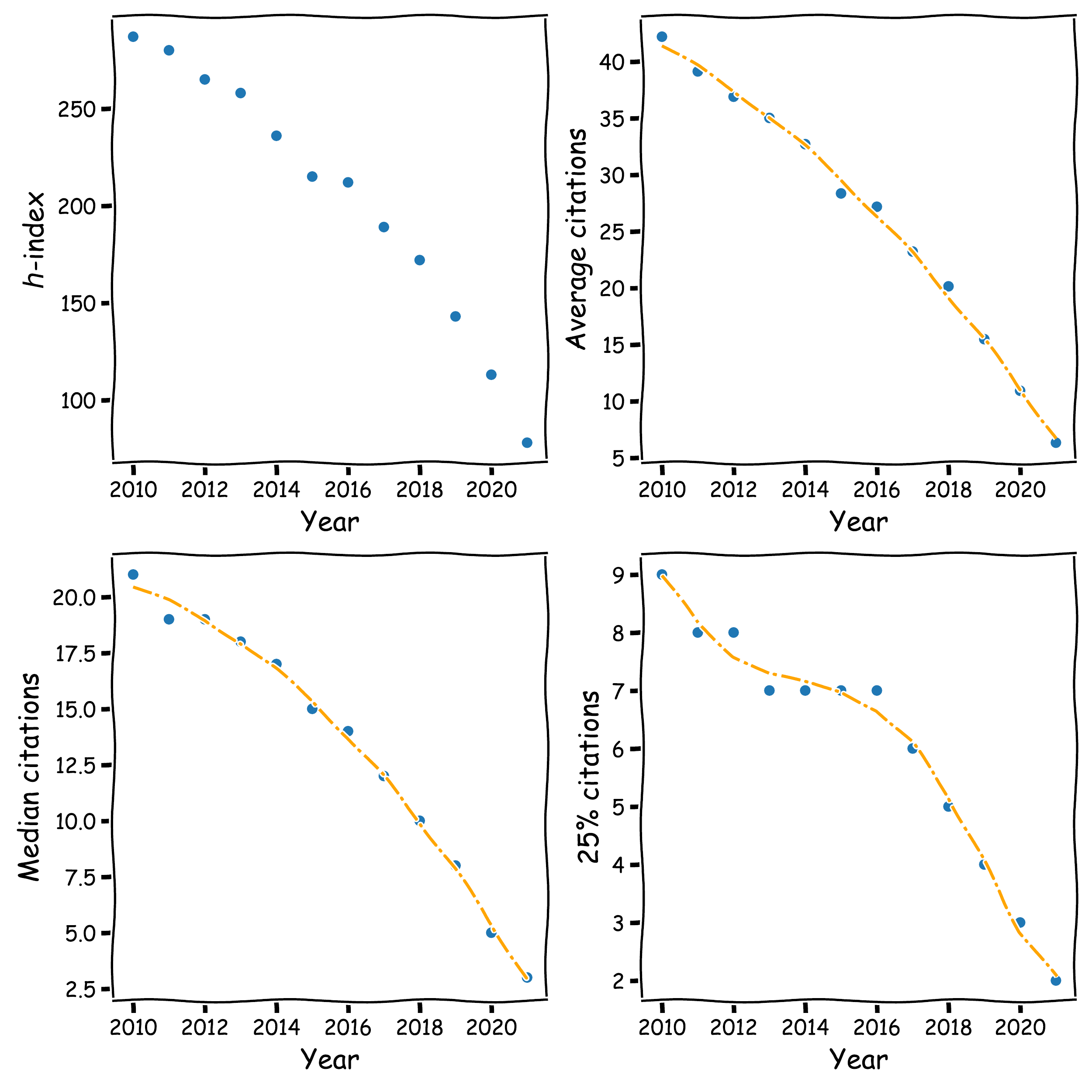}
    \caption{The $h$-index of all astronomical related refereed papers from the NASA ADS, published in a given year (top left) and the average, median and 25\% quartile of the number of citations. The dash-dotted line is a tentative fit.}
    \label{fig:ads}
    	\end{center}
\end{figure}

\section{As in every good game, let's have a penalty}
If the above scheme is already remarkable\footnote{And therefore unlikely to be original.}, it fails to address another of the syndromes affecting current science: the pollution of the literature. Refereed journals\footnote{Not to mention in predatory journals!} host indeed too many papers that do not really increase significantly Humanity's corpus of knowledge. This is particularly true in astronomy it seems, as most journals boast a rejection rate that is between 5 and 10\%. This may make authors rather happy, but does it make a service to the community? As a discussion about this is only worth having over a beer and will likely be as useful as to decide which is the best parenting mode (not too lenient, not too strict, where is the middle?), I'm not even going to consider making the beginning of an embryonic start to an iota of an argument\footnote{But readers should feel encouraged to invite me for a beer to discuss this!}. Let me simply state that we should be more drastic about this. Thus, I propose that a penalty function is introduced when calculating the $b$-index. 
In the typical manner that characterises the methodology presented in this paper, such a penalty function should of course be based on some firm ground. 

One could (and one certainly will) argue that if a paper is not cited, the citation index of an author doesn't increase and there is no need to do anything else. This may be true, but this is too simple a reasoning. In the same way as the $b$-index derived above will discourage people from adding unnecessary co-authors on their paper, my aim is to deter people from publishing non-essential refereed papers. If this makes me appear as a Don Quixote battling with winesacks, so be it.

To look at the possible penalty, I looked at all the yearly refereed papers between 2010 and 2021, and their citations, as reported in the NASA ADS. Some graphical representation of this is shown in Fig.~\ref{fig:ads}. One of the panels shows the evolution of the $h$-index of this set of papers, revealing how this is dependent on the scientific age. 

Looking as an example at the refereed papers published in 2010, that is, quite old enough to have gathered many citations in principle, I get 24,572 papers that collected 917,390 citations. Of these, 2\,829 didn't get a single citation, but this is partly due to the imprecise way that the ADS selects refereed papers. In the following, I therefore took the conservative approach to ignore these papers with no citation. A further 1\,104 have only one citation (after 12 years), so it is perhaps justified to consider if they should at all have been published. It is not for me to comment on why papers dealing with multifractality of short gamma-ray bursts or on the jovian Trojan dust do not get cited, but this is apparently another sad fact of life that shouldn't make us despair, however\footnote{The reader shouldn't think that I consider myself better than the rest of us. I do have two refereed papers that have gathered exactly zero citations since they exist! While this is perhaps telling much more about me -- could I really never find a reason to cite these? -- than about science itself, one can only feel saddened, in the same way that one sees our preferred artist finishing last at the Eurovision Song Contest! Perhaps also a referee or an editor could have been more honest and tell me that this was a waste of time and paper? (Articles were indeed still printed on paper at the time.)}. 

Figure~\ref{fig:ads} also shows the average and median number of citations of all the refereed papers (minus those without any citation) published in a given year and listed by ADS. The average is skewed because of the few papers that have many citations. In 2010, the most cited paper has almost 5\,000 citations! But as shown by the $h$-index, the 287th most cited paper of that year has ``only'' 287 citations. Perhaps the median is more representative and is also shown in the figure. It can be seen that it increases for older papers, following the law:
$$
\tilde{C} = -0.096*x^2 + 1.383 * x +  16.229,
$$
where $x=$year$-2000$. Using the median is perhaps quite harsh, as we do not want to penalise half of all published papers! 

So, instead, let us look at the final panel of Fig.~\ref{fig:ads}, which displays the first quartile of the citation's distribution. Thus, in 2010, the 25\% less cited papers had less than 9 citations, a number that went down to 7 in 2016, and then linearly decreased by one unity every year since then. The evolution of this quartile is clearly very complicated, showing a plateau between 2013 and 2016, and even if I can reproduce it with a 5th order polynomial, this is likely overfitting. Instead, I propose a less cumbersome and more generous approach, using a linear increase for the five years following the publication, and a saturation at 7 citations after that. That is, we can define a minimum number of citations, $Q_{\rm min}$:
\[
    Q_{\rm min} = 
\begin{cases}
    2+y,& \text{if } y=\text{now - 1 - pubyear} \leq 5\\
    7              & \text{otherwise.}
\end{cases}
\]

If a paper is published in a given {\it pubyear} has {\it now} less than $Q_{\rm min}$ citations, a penalty should be applied, that is, the effective number of citations of this paper becomes $C_{\rm eff}=C - \beta (Q_{\rm min}-C)$, with a possible value of $\beta$ being $0.2 N$, as one should assume that the more authors there are, the more cited a paper should be, if we want not to waste resources. 
This value of $C_{\rm eff}$, which can thus be negative in some cases, should then be used in the definition of the $b$-index. 

\section{Famous last words}
The above methodology presents, hopefully in a pellucid prose, a new way to compute the impact and productivity of a scientist, avoiding the many biases carried by the notorious and ubiquitous $h$-index. Once the community has been convinced of the drastic improvements such a new index represents, they should adjust their publishing practices and provide the necessary information about the authors such that the index can be easily computed. This being said, it is also important that any further decision being taken should take into account, among others, the whole career and its interruptions, the gradient of the $b$-index, the originality of the ideas, etc.

\vspace{0.25cm}
{\it --``Well, this becomes so complicated that perhaps we should simply look at the CV and the quality of the papers published?''

-- You may be right! Perhaps this is what you should do!}

\vspace{0.25cm}
You are not a number. Be seeing you.

\vspace{0.5cm}
\begin{acknowledgements}
As for the  other  pieces in this series, this work was done outside of working hours, when the first author was bothered by his annoying colleagues, A. Wake, W.H.Y. Can't, I. Sleep. 

This research has made use of NASA’s Astrophysics Data System Bibliographic Services.

Matplotlib \citep{2007CSE.....9...90H} and its {\it xkcd} environment are gratefully acknowledged.
\end{acknowledgements}

\end{document}